\documentclass[fleqn,10pt]{wlscirep}
\usepackage[utf8]{inputenc}
\usepackage[T1]{fontenc}
\usepackage{multirow}
\usepackage{graphicx}
\usepackage{subfigure}

\title{Self-Attentive Sequential Recommendation with Cheap Causal Convolutions}

\author[1]{Jiayi Chen}
\author[1,2,*]{Wen Wu}
\author[1]{Liye Shi}
\author[1]{Yu Ji}
\author[3]{Wenxin Hu}
\author[2]{Xi Chen}
\author[4]{Wei Zheng}
\author[1]{Liang He}
\affil[1]{School of Computer Science and Technology, East China Normal University}
\affil[2]{Shanghai Key Laboratory of Mental Health and Psychological Crisis Intervention, School of Psychology and Cognitive Science, East China Normal University}
\affil[3]{School of Data Science and Engineering, East China Normal University}
\affil[4]{Information Technology Services, East China Normal University}
\affil[*]{wwu@cc.ecnu.edu.cn}


\begin{abstract}
Sequential Recommendation is a prominent topic in current research, which uses user behavior sequence as an input to predict future behavior. By assessing the correlation strength of historical behavior through the dot product, the model based on the self-attention mechanism can capture the long-term preference of the sequence. However, it has two limitations. On the one hand, it does not effectively utilize the items' local context information when determining the attention and creating the sequence representation. On the other hand, the convolution and linear layers often contain redundant information, which limits the ability to encode sequences. In this paper, we propose a self-attentive sequential recommendation model based on cheap causal convolution. It utilizes causal convolutions to capture items' local information for calculating attention and generating sequence embedding. It also uses cheap convolutions to improve the representations by lightweight structure. We evaluate the effectiveness of the proposed model in terms of both accurate and calibrated sequential recommendation. Experiments on benchmark datasets show that the proposed model can perform better in single- and multi-objective recommendation scenarios.
\end{abstract}
\begin{document}

\flushbottom
\maketitle
%
%
\thispagestyle{empty}

\section{Introduction}
Recommendation systems have become a research hotspot in academia and industry by analyzing users' historical behaviors and actively providing users with content that may be of interest from a variety of resources. Early recommendation models can be mainly classified into collaborative filtering models \cite{qiu22duorec,xie22cl4sr}, content-based models \cite{pazzani07contentrs,lops11contentrssota}, and hybrid recommendation models combining the two \cite{kim06combining,phuong08graphhybrid}. Sequential recommendation predicts items that users may interact with in the future based on users’ behavior sequences. It has become an important direction of recommender systems in recent years. With the development of deep learning, researchers have used deep neural networks to model behavioral sequences and made progress in accurately predicting users' future behaviors.
Among various structures of sequence modeling, self-attention has been extensively used in sequential recommendation \cite{kang18sasrec,qiu22duorec,xie22cl4sr,xu21longshortsa,li21lightweight,sun19bert4rec,moreira21transformer4rec}. 

Sequential Recommendation based on the self-attention mechanism can capture long-term preferences of sequences by calculating correlations among historical items. However, the dot product for calculating attention has limitations in perceiving items’ contextual information. It only utilizes the embeddings of corresponding items to compute the correlation, which assumes items are independent of adjacent items. However, items are often related to their contextual items, and incorporating these items can improve the representations of items. 
Similarly, existing models usually predict future behavior based on the last item of the sequence, which also neglects the last item's local information. 
In addition, linear and convolution layers are used in sequential recommendation, such as generating sequence embeddings or gathering contextual information. However, they usually contain redundant information\cite{han20ghostnet}, because feature maps often contain similar information, which limits the ability to encode sequences. These layers also require a large parameter size. Therefore, it is necessary to design a lightweight mechanism to generate richer information.

To address the above problems, we propose a self-attention sequential recommendation model with cheap causal convolutions, namely C3SASR. C3SASR first uses causal convolution \cite{bai21causalconv} to fuse the contextual information. The causal convolution takes the current item and some of its predecessors as input, obtains a representation of the item fused with the sequence local information through the convolution layer, and subsequently inputs the representation of the item to the self-attentive layer for the computation of attention and the generation of the item representation. In generating the sequence representation, the exact causal convolution mechanism is used to fuse the local information according to the last item of the sequence. The information of the item itself is weighted and summed to obtain the final sequence representation. For the redundant information in the convolutional and linear layers, we design a cheap convolutional mechanism. It first takes the item representation as the input and reduces the number of convolution channels by half to obtain a representation with half of the original dimension. Then the reduced-dimensional representation is linearly processed, enhanced, and spliced with the reduced-dimensional representation to obtain a new representation with the same dimension as the input representation. Compared with the ordinary linear and convolutional layers, our model architecture can obtain a richer representation and is more lightweight. 
We evaluate the effectiveness of the proposed model in both single- and multi-objective recommendation scenarios. The single-objective recommendation only focuses on the recommendation accuracy, and the multi-objective recommendation concentrates on calibrated recommendation\cite{steck18calibrated} along with accuracy. Experiments show that the C3SASR model can effectively improve the model's ability to encode sequences and achieve better performance.

The specific contributions of this paper are as follows:
\begin{itemize}
    \item[(1)] In this paper, we improve the sequence encoder for the self-attentive mechanism so that the model has better sequence encoding capability to better serve the calibration and accurate optimization problems with constraints.
    \item[(2)] We merge the contextual information using causal convolution to address the point-by-point calculation issue of the self-attentive mechanism. Additionally, the model uses causal convolution to improve the local information of the final item in the sequence while creating sequence representations, better capturing the function of short-term preferences.
    \item[(3)] We employ a cheap convolution method that can produce better item and sequence representations at a lighter parameter scale to address the issues of the numerous parameters and redundant information in the linear and convolutional layers.
    \item[(4)] Experiments on common datasets show that C3SASR can improve performance of both accurate and calibrated recommendations.
\end{itemize}

\section{Related Work}
In this section, we do a literature review of our related work. We first introduce the recent advances in sequential recommendation. Next, we briefly introduce the calibrated recommendation. Finally, we also review the advanced convolutions.

\subsection{Sequential Recommendation}
Hidasi et al. \cite{hidasi2015session}first used Gated Recurrent Units (GRU) \cite{cho14gru} for Sequential Recommendation, proposing to input sequences into GRU to get a representation of the sequence and get predicted scores for all items by feedforward neural networks. Li et al. \cite{li2017neural} designed attention based on the sequence representation obtained by the GRU mechanism to capture the main purpose of the sequence. Ren et al. \cite{ren2018repeatnet} introduced the repetition exploration mechanism into the GRU-based sequence encoder to predict when to make repetitive recommendations after considering the user's repetitive behavior in the sequence. On the other hand, graph neural network-based algorithms transform sequences into graph structures and thus learn transfer relations between items \cite{wu2018session,qiu19rethinking,pan20stargraph,xu19graphcontext,chang21seqgnn,gcegnn,xia21ssgraphcotrain,zheng20dgtn}. Wu et al. \cite{wu2018session} presented the SR-GNN model, which converts sequences into directed graphs, learns the representation of items through gated graph neural networks, and the representation of sequences using an attention mechanism. Pan et al. \cite{pan20stargraph} converted sessions into graph structure by adding a virtual node to learn transfer relations between long-spaced items. Wang et al. \cite{gcegnn} built session-level and global-level graph networks to improve recommendation performance by introducing information from other sessions. Xia et al. \cite{xia21ssgraphcotrain} designed graph structures from two perspectives, items, and sessions, respectively, and trained the network using self-supervision and contrastive learning to obtain better recommendation accuracy. The SASRec model \cite{kang18sasrec} used a self-attention mechanism to learn the long-term and short-term preferences of user behavior sequences and achieved satisfactory recommendation accuracy. Xu et al. \cite{xu21longshortsa} divided sequences into subsequences and captured users’ long-term and short-term preferences by applying two self-attention networks. Long-term and short-term preferences. Li et al. \cite{li21lightweight} devised a lightweight sequential recommendation model based on the self-attentive mechanism, which reduces the number of parameters in the item representation layer.

\subsection{Calibrated Recommendation}
The concept of recommendation calibration originates from the field of machine learning. In the field of recommender systems, Steck \cite{steck18calibrated} first suggested calibration to reduce the difference between the recommendation results and the preference distribution of the user's historical behavior. According to the classification system of fairness criteria \cite{burke17multisidefair}, calibration belongs to user-level fairness \cite{steck18calibrated, silva21exploiting}. Its fairness is reflected by narrowing the preference distribution between recommendation lists and user behavior. Steck et al. \cite{steck18calibrated} constructed a calibration-oriented reordering model. The model is based on a greedy strategy that considers the gain of adding candidate items to accuracy and calibration at each step of generating the recommendation list, controlling the status of accuracy and calibration through a weight parameter. Abdollahpouri et al. \cite{himan20connection} investigated the link between popularity bias and calibration, where users who are more influenced by popularity bias tend to get uncalibrated recommendation lists. Kaya et al. \cite{Kaya19comparison} compared a diversity-oriented intention-aware algorithm with a calibration-oriented reordering algorithm to make recommendation lists more diverse. Seymen et al. \cite{seymen21constrain} created a new way to measure the consistency of preference distributions for recommendation lists and historical behaviors, thus improving the calibration performance of the reordering algorithm. Nazari et al. \cite{nazari22choice} targeted the podcast scenario by introducing a calibrated recommendation algorithm that can meet the diverse needs of users. A weighted loss function is used in DACSR \cite{dacsr} to trade off calibration and accuracy, and calibrated sequential recommendation becomes an optimization problem with constraints.

\subsection{Advanced Convolutional Architectures}
Yuan et al. \cite{yuan19nextitnet} constructed a simple but very effective recommendation model with the ability to learn advanced representations from short- and long-term item dependencies. The network structure of the model consists of a series of perforated convolutional layers that can effectively increase the perceptual field without relying on pooling operations. BAI et al. \cite{bai18cnnrnn} systematically evaluated generic convolutional and recursive architectures for sequence modeling. The results showed that simple convolutional structures outperformed standard recursive networks (e.g., LSTM) across various tasks and datasets while showing longer adequate memory. Yang et al. \cite{yang22lightweight} proposed an efficient feature-cheap convolutional super-resolution (FCCSR) model, which consists of multiple multilevel information fusion blocks (MIFBs). In addition, FCCSR designs a novel lightweight channel attention module for obtaining fewer parameters and better performance. It provides a practical idea for designing lightweight recommender systems. Further development of the cheap convolution mechanism \cite{han20ghostnet,elliot18moonshine} allowed us to build efficient and concise recommendation models. 

For the self-attentive sequential recommendation, local information of items is not well used. Therefore, we utilize causal convolutions to incorporate local information before the computation of attention. Either aggregation of local item information or generation sequence embeddings needs convolution and sequential layers. For the redundant information and huge parameter scale of these layers, we utilize cheap convolutions to obtain better representations with a lightweight structure. 

\section{Methods}
\begin{figure}
    \centering
    \includegraphics[width=15cm]{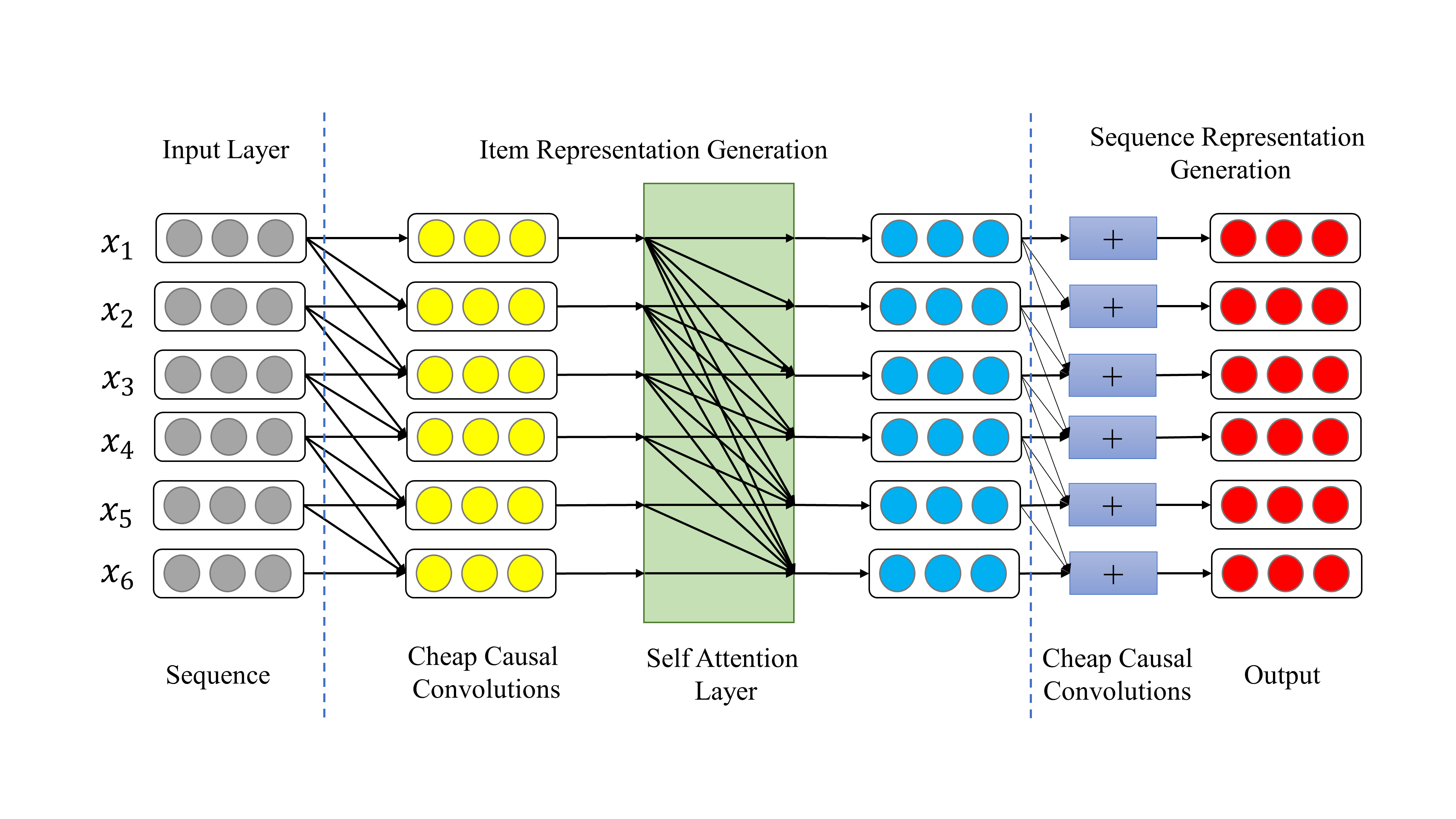}
    \caption{The architecture of C3SASR model.}
    \label{fig:c3sasr}
\end{figure}

In this section, we introduce the proposed C3SASR series model. Although there are various sequential recommendation models, the commonplace approaches have a layered structure. These models typically have four layers: input layer, item representation generation layer, sequence representation generation layer, and prediction layer. Our model modifies the item representation generation layer and sequence representation generation layer. Fig. \ref{fig:c3sasr} depicts the overall structure of the C3SASR model.

\subsection{Causal Convolutions}
\begin{figure}[tbp]
\centering
\subfigure[Generation of item representation]{
\includegraphics[width=8cm]{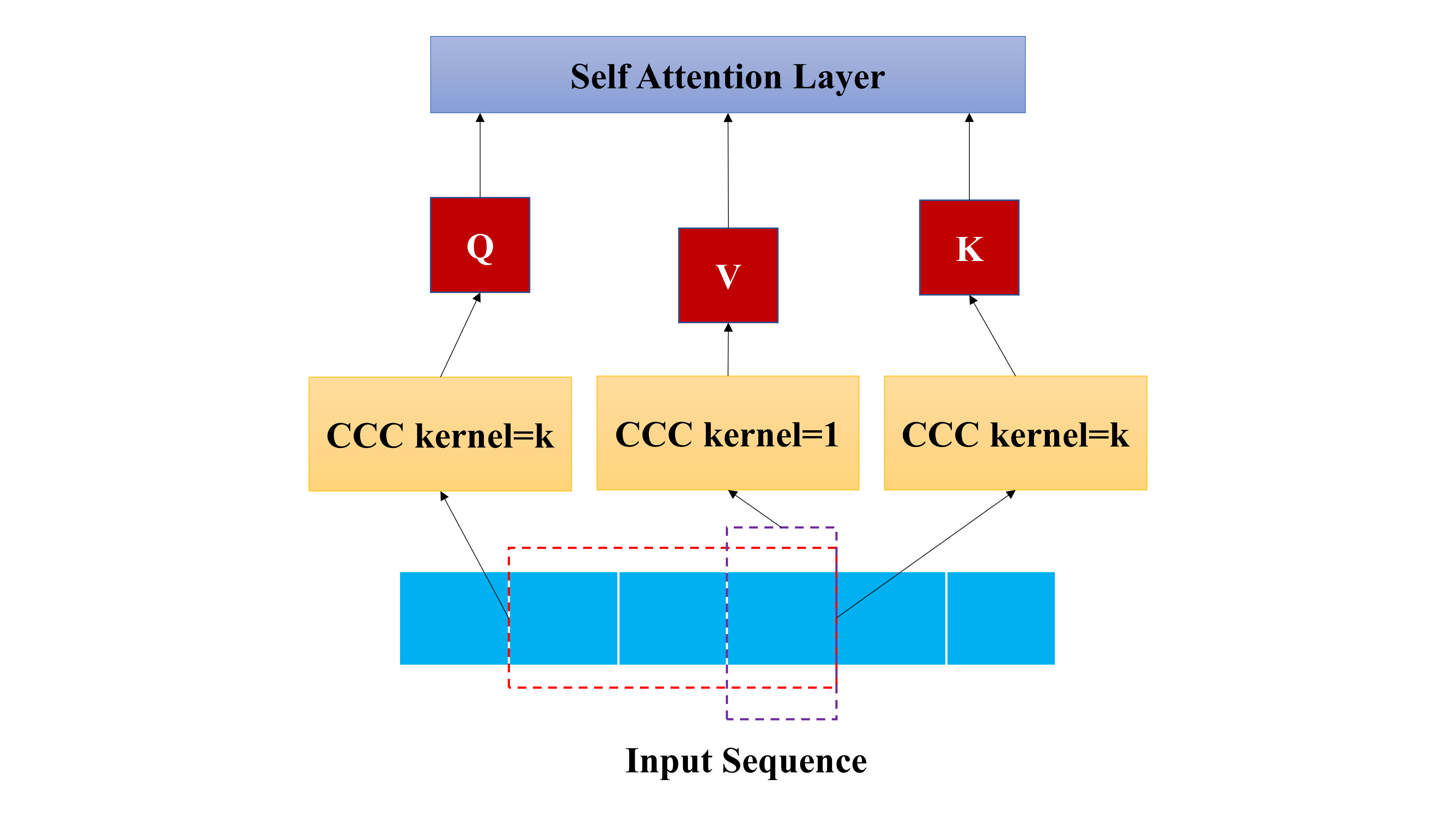}
\label{subfig:c3sasr:sa}
}
\quad
\subfigure[Generation of sequence embedding]{
\includegraphics[width=8cm]{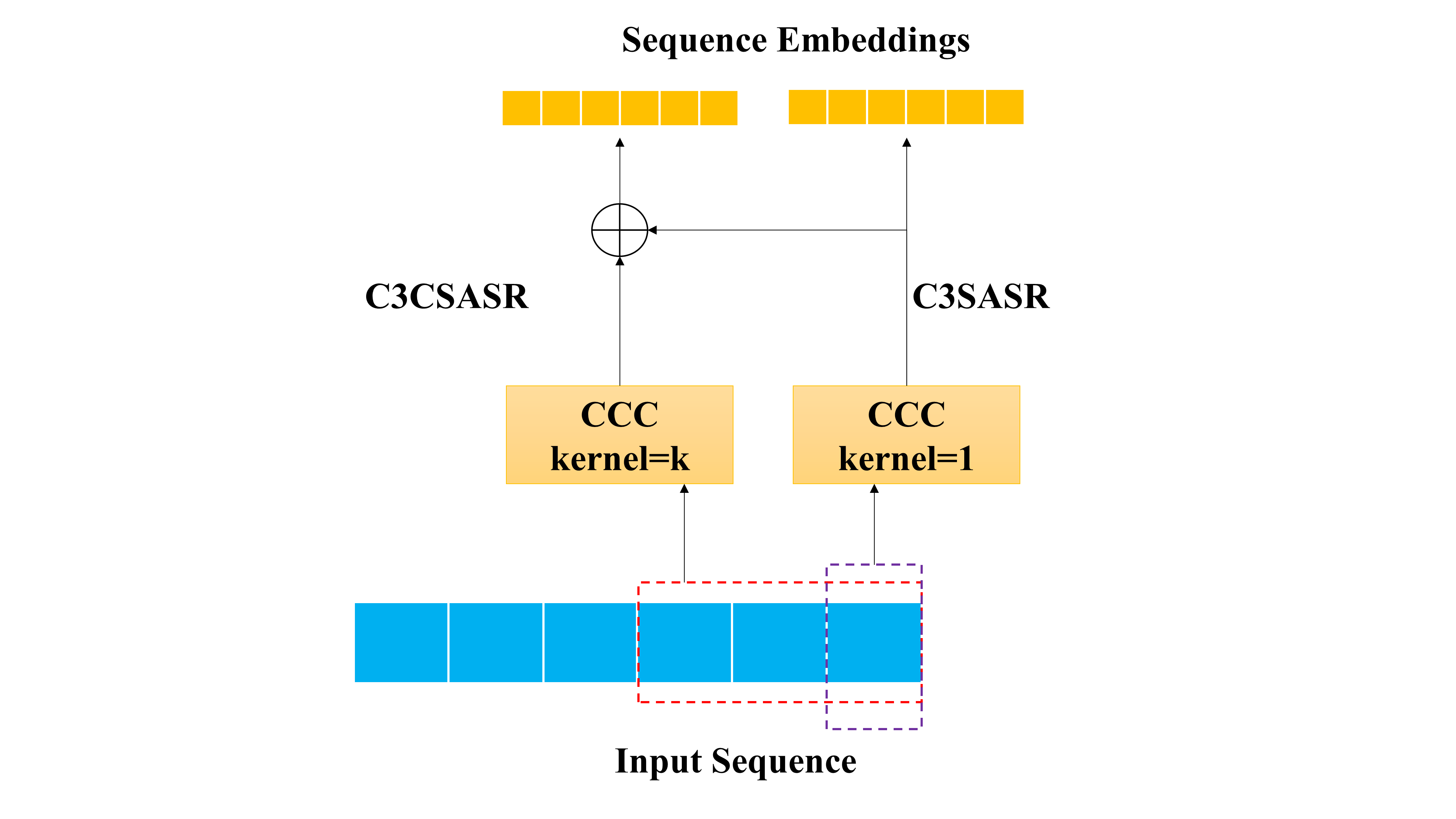}
\label{subfig:c3sasr:out}
}
\caption{The architecture of item and sequence representation generation.}
\label{fig:c3sasr:local}
\end{figure}

The process of creating item and sequence representations now includes a local augmentation method based on cheap causal convolution. In Fig. \ref{fig:c3sasr:local}, the precise structure is displayed. Firstly, after obtaining the corresponding representations of sequential items in the input layer, item $x_t$ is formed into a k×d-dimensional input with the previous k-1 items. The process can be written as
\begin{equation}\label{eq:1}
h_{x_t} = CCConv([x_{t-k+1},..., x_t])
\end{equation}
where $CCConv(·)$ is the proposed cheap causal convolution (C3) mechanism. $h_{x_t}$ is the representation of item $x_t$ obtained after the C3 mechanism. d is the dimension of the item representation. 
The self-attentive mechanism changes the item representations into Query (Q), Key (K), and Value (V) by linear transformation, respectively. Similarly, we use different causal convolutional networks to fuse local information. For Query and Key, a causal convolution mechanism of length k is used as:
\begin{align}
Q &= W_Q CCConv_Q([x_{t-k+1},...,x_{t-1}, x_{t}])\\
K &= W_K CCConv_K([x_{t-k+1},...,x_{t-1}, x_{t}])
\end{align}
where $CCConv_Q$ and $CCConv_K$ are the causal convolution modules for Q and K, and WQ and WK are the linear transformation layers employed in the self-attentive mechanism. 
Furthermore, V uses the causal convolution mechanism with k=1, which is expressed as:
\begin{equation}
V = W_V CCConv_V([x_t])
\end{equation}
At this point, the causal convolution degenerates to transform only for the representation of item $x_t$, without considering the information of the context. After obtaining the Query, Key, and Value, the representation of each item in the sequence is obtained according to the algorithm of the self-attentive mechanism as:
\begin{equation}
h^{sa} = softmax(\frac{QK^\mathsf{T}}{\sqrt{d}})V
\end{equation}
where $h^{sa}=\{h_{x_1}^{sa},h_{x_2}^{sa},...,h_{x_t}^{sa}\}$ is the representation of all items in the sequence obtained after the self-attentive layer, and d is the dimensionality of the items and the sequence representation.

Similarly, in generating the sequence representations, this paper also proposes to enhance the nonlinear capability of the model using a locally enhanced cheap convolution mechanism. Firstly, after obtaining the item representations through the self-attentive layer, the model uses a cheap convolution mechanism to transform the output of the self-attentive layer. The process can be written as:
\begin{equation}
o^{self}_{x_t} = CCConv(ReLU(CCConv([h_{x_t}^{sa}])))
\end{equation}
where$h_{x_t}^{sa}$ represents the item$x_t$ obtained after a self-attentive layer, which is used as input of the cheap convolution module. $o^{self}_{x_t}$ is the representation of item $x_t$ after a cheap convolution mechanism. It is similar to the feedforward network structure of the SASRec model. The difference is that the C3SASR model uses cheap convolution as the major transformation.

In addition, models such as SASRec use the representation of the last item in the sequence as the sequence representation. In this paper, we introduce the last item's local information to enhance the sequence's representation. Similarly, a light causal convolution mechanism is used to obtain information about the current item context as:
\begin{equation}
o^{cxt}_{x_t} = CCConv(ReLU(CCConv([h_{x_{t-k_{2}+1}}^{sa},...,h_{x_t}^{sa}])))
\end{equation}
where$k_2$ is the range of contexts considered in the sequence representation generation layer and also the convolution kernel’s size. Ultimately, the representation of item $x_t$ can be obtained by weighting its representation $o^{self}_{x_t}$ with the contextual representation $o^{cxt}_{x_t}$ as
\begin{equation}
o^w_{x_t} = \alpha o^{self}_{x_t} + (1 - \alpha)o^{cxt}_{x_t}
\end{equation}
where $\alpha$ represents the weighted factor. $o^w_{x_t}$ is the weighted representation of item $x_t$. 

In this paper, the model that uses its representation $o^{self}_{x_t}$ as the final item representation $o_{x_t}$ of item $x_t$ is abbreviated as C3SASR, while the model that uses the weighted representation $o^w_{x_t}$ as $o_{x_t}$ is abbreviated as C3CSASR.

\subsection{Cheap Causal Convolutions}
\begin{figure}
    \centering
    \includegraphics[width=12cm]{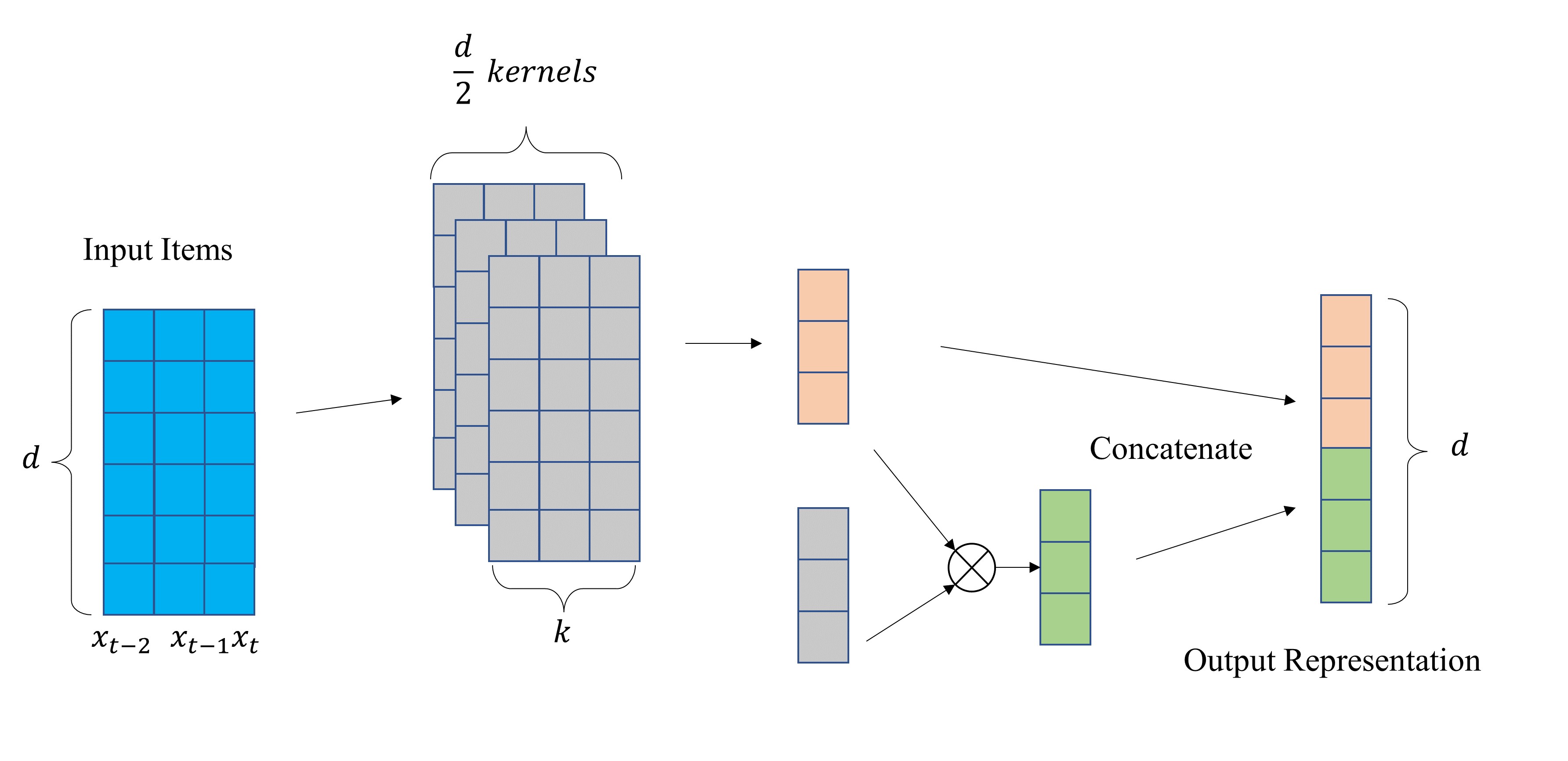}
    \caption{The structure of cheap convolution}
    \label{fig:cheap}
\end{figure}
In this section, we detail the cheap convolution mechanism employed in causal convolution (Eq. \ref{eq:1}), whose structure is shown in Fig. \ref{fig:cheap}. 
This mechanism encodes the input $k \times d$-dimensional short-term sequence representation into a $d$-dimensional vector representation in a convolutional manner. 
It first compresses the input $k \times d$-dimensional representation into a $d$-dimensional representation. 
Its employs $d/2$ convolution kernels $\{c_1,c_2,...,c_{d/2}\}$, each with a $k \times d$ dimensional convolution kernel. 
The specific procedure for generating vector representations is given as 
\begin{equation}
e^c_{x_t} = [Sum(c_1, e_t), Sum(c_2, e_t),..., Sum(c_{d/2}, e_t)]
\end{equation}
\begin{equation}
Sum(c_l, e_t) = \sum_i \sum_j c_l[i, j] \cdot e_t[i, j]
\end{equation}
where $e_t=[x_{t-k+1},...,x_{t-1},x_t]$ represents a local sequence representation of the input. $Sum(c_l,e_t)$ operation represents the convolution kernel $c_l$ multiplied with the corresponding elements of the input representation $e_t$ and summed. The output $e_{x_t}^c$, on the other hand, is stitched together from the outputs of d/2 convolution kernels. To obtain a deeper representation of the sequence, the output $e_{x_t}^c$ is augmented as
\begin{align}
e^{c2}_{x_t} &= e^c_{x_t} \odot W_c \\
e_{x_t} &= Tanh([e^c_{x_t}, e^{c2}_{x_t}])
\end{align}
The dimension of $W_c$ is the same as that of $e_{x_t}^c$, which is the parameter to be learned. $e_{x_t}^c$ will be multiplied with its corresponding element to obtain an enhanced representation and spliced with itself to obtain the final representation $e_{x_t}$. Compared with traditional convolution, the cheap convolution mechanism requires $k \times d \times d/2 + d/2$ parameters. The dimension $d$ usually takes a more considerable value (e.g., 64 or 100), so that the number of parameters will also be substantially smaller than the k×d×d parameters of the traditional 1D convolution.

\section{Model Training and Prediction}
After obtaining the final representation $o_{x_t}$ of item $x_t$, the score of $x_j$ is obtained by making a dot product of $o_{x_t}$ with the item embedding $E_j$ and normalized by the softmax function:
\begin{equation}
\hat{y}_j = softmax(E_j \cdot o_{x_t})
\end{equation}
where $\hat{y}_j$ is the predicted score of item $j$. 

In this paper, we aim to evaluate the ability to encode sequences under both single and multiple objectives, and choose the emerging concept ``Calibration'' as another objective in addition to accuracy. 
We adopt the weighted sum of loss functions based on accuracy and calibration as
\begin{equation}
L_w = (1 - \lambda) \times L_{Acc}(y, \hat{y}) + \lambda \times L_{Calib}(\hat{y})
\end{equation}
where $L_{Acc}$ and $L_{Calib}$ are loss functions for accuracy and calibration, respectively. By the weighted loss function, we can evaluate the effectiveness of single or multiple objectives of our model. By setting $\lambda$ to 0, the model is only optimized toward accuracy. And the $\lambda > 0$ means the model is dealing with both accuracy and calibration.

$L_{Acc}$ is the accuracy-based loss function, and the cross entropy is used as the loss function in this paper:
\begin{equation}
    L_{Acc} = \sum_{i=1}^{\mid I \mid} y_i \log \hat{y}_i
\end{equation}
where $y_i=1$ means item $i$ is the user's next item, and 0 other wise. $\mid I \mid$ is the number of all items. 
$L_{Calib}$ is the calibration based loss function \cite{dacsr}. 
Following \cite{dacsr}, the loss function for calibration can be written as: 
\begin{equation}
    L_{Calib}(\hat{y}) = 1 - cos(\hat{q}(s), p(s))
\end{equation}
which measures the consistency of the preference distributions between recommendations $\hat{q}(s)$ and historical behaviors $p(s)$. The calculation process can be found in \cite{dacsr}.

\section{Experiments}

\subsection{Dataset}
We adopted two commonly-used benchmark datasets to evaluate the performance of our model. The first one is \emph{Movelens-1m}(https://grouplens.org/datasets/movielens/1m), which contains interaction logs of more than 6000 users and 3000 movies. The other is \emph{Tmall}(https://tianchi.aliyun.com/dataset/dataDetail?dataId=53), which includes user behavior logs on an e-commerce platform. We follow the previous work\cite{dacsr} to process the datasets. The statics of datasets are listed in Table. \ref{table-statistic}.
\begin{table}[t]
\centering
\caption{Statistics of Datasets}
\begin{tabular}{cccc}
    \hline
     Statistics  & Ml-1m & Tmall\\
     \hline
     Number of users & 6,040 & 31,854\\
     Number of Items  & 3,883 & 58,343 \\
     Number of Training Sequences & 981,504 & 832,603  \\
     Number of Testing Sequences  & 6,040 & 31,854 \\
     Number of Attributes & 18 & 70 \\
     Average Length of Sequence & 164.50 & 28.13 \\
     \hline
\end{tabular}
\label{table-statistic}
\end{table}

\subsection{Evaluation Metrics}
We evaluate models from accuracy and calibration perspectives. Following previous work\cite{wu2018session, li2017neural}, we use Recall and MRR as evaluation metrics to measure the recommendation accuracy of models. 
\begin{itemize}
    \item \textbf{Recall@K} (Rec@K) is a widely used metric in recommendation and information retrieval areas. Recall@K computes the proportion of correct items in the top-K items of the list.
    \begin{equation}
        Recall@K= \frac{1}{N} \sum_{s} 1(x_{n+1} \in RL_{s})
    \end{equation}
    where $1(\cdot)$ is an indication function whose value equals 1 when the condition in brackets is satisfied and 0 otherwise. $N$ is the number of testing cases.
    \item \textbf{MRR@K} is another important metric that considers the rank of correct items. The score is computed by the reciprocal rank when the rank is within K; otherwise the score is 0. 
    \begin{equation}
        MRR@K = \frac{1}{N} \sum_{s} \frac{1}{rank(x_{n+1}, RL_{s})}
    \end{equation}
\end{itemize}

In addition to accuracy, we adopt $C_{KL}$ \cite{steck18calibrated} metric to evaluate the performance of calibration. The $C_{KL}$ compares the consistency between two preference distributions:
\begin{equation}
    C_{KL}(RL, s) = \frac{1}{N}\sum_{s} KL(q(s) \mid p(s))
\end{equation}
where $KL(\mid)$ is the Kullback-Leibler Divergence. The lower value of $C_{KL}$ means more calibrated recommendations are generated. The preference distribution $p(s)$ and $q(s)$ represent distributions of user's historical behaviors and generated recommendations, respectively. The distribution is measured by item attributes (e.g., movie genres), which is same to previous work\cite{steck18calibrated}. 

\subsection{Experiment Setup}
The dimensionality of the object and sequence representations is set to 64. the convolution kernel size $k$ is set to 3 for the C3 mechanism of our C3SASR, $k_2$ is set to 3 for the C3CSASR model, and $\alpha$ is set to 0.8. Adam [29] is used as the optimizer, the learning rate is set to 0.001, and the batch sizes on the Ml-1m and Tmall datasets are 256 and 128. We still measure the performance of the model in terms of both accuracy and calibration. In this paper, we use three metrics, Rec@K, MRR@K, and $C_{KL}@K$, where K is taken as 10 and 20.

\subsection{Baselines}
We use the following models as a comparison: 
\begin{itemize}
    \item \textbf{SASRec} \cite{kang18sasrec} is a sequential recommendation model based on self-attention, which has achieved satisfactory recommendation accuracy in existing research. It is a powerful baseline model. 
    \item \textbf{CaliRec} \cite{steck18calibrated} is a reprocessing model based on post-processing, which reorders the candidate list generated by the recommendation model. It uses a sorting mechanism based on greed to balance the relationship between accuracy and calibration through a super parameter.
    \item \textbf{DACSR} \cite{dacsr} is a decomposition aggregation framework that uses the SASRec model as the sequence encoder.
\end{itemize}

\section{Result and Analysis}
In this section, we aim to answer the following research questions:
\begin{itemize}
    \item RQ1: Whether our proposed C3SASR model can outperform the baselines?
    \item RQ2: How the performances change when the hyper-parameters change?
    \item RQ3: How does each module contribute to the performance?
\end{itemize}

\subsection{RQ1: Overall Performance}
In this section, we give the performance comparison between the C3SASR model and the baseline model. We evaluate the performances of our model and baselines under single- and multiple- objective scenario. We first focus on the performance of single sequence encoder, then we conduct experiments under the DACSR framework \cite{dacsr}. 


\begin{table}[tbp]
  \centering
  \caption{Performance comparisons between C3SASR and SASRec}
    \begin{tabular}{c|c|c|cccccc}
    \toprule
    Dataset & Models & $\lambda$ & Rec@10 & MRR@10 & $C_{KL}$@10 & Rec@20 & MRR@20 & $C_{KL}$@20 \\
    \midrule
            \multirow{6}[3]{*}{Tmall} & SASRec & \multirow{3}[1]{*}{0} & 0.1512  & 0.0847  & 2.4871  & 0.1860  & 0.0871  & 2.1092  \\
          & C3SASR &       & \textbf{0.1549 } & \textbf{0.0869 } & 2.5430  & \textbf{0.1876 } & \textbf{0.0891 } & 2.1709  \\
          & C3CSASR &       & 0.1526  & \textbf{0.0869 } & \textbf{2.4703 } & 0.1854  & \textbf{0.0891 } & \textbf{2.0962 } \\
\cmidrule{2-9}          & SASRec & \multirow{3}[2]{*}{0.5} & 0.1508  & 0.0854  & \textbf{2.1490 } & 0.1830  & 0.0876  & \textbf{1.7810 } \\
          & C3SASR &       & 0.1498  & 0.0868  & 2.2005  & 0.1828  & 0.0890  & 1.8376  \\
          & C3CSASR &       & \textbf{0.1532 } & \textbf{0.0891 } & 2.2168  & \textbf{0.1876 } & \textbf{0.0915 } & 1.8495  \\
    \midrule
    \multirow{6}[4]{*}{Ml-1m} & SASRec & \multirow{3}[2]{*}{0} & 0.2627  & 0.1102  & 1.3029  & 0.3675  & 0.1174  & 0.9045  \\
          & C3SASR &       & 0.2624  & 0.1120  & 1.2855  & 0.3659  & 0.1190  & 0.9032  \\
          & C3CSASR &       & \textbf{0.2717 } & \textbf{0.1193 } & \textbf{1.2618 } & \textbf{0.3738 } & \textbf{0.1263 } & \textbf{0.8772 } \\
\cmidrule{2-9}          & SASRec & \multirow{3}[2]{*}{0.5} & 0.2551  & 0.1088  & \textbf{1.0780 } & 0.3599  & 0.1161  & \textbf{0.7281 } \\
          & C3SASR &       & 0.2593  & 0.1118  & 1.0849  & \textbf{0.3707 } & \textbf{0.1195 } & 0.7390  \\
          & C3CSASR &       & \textbf{0.2606 } & \textbf{0.1120 } & 1.0987  & 0.3674  & 0.1194  & 0.7534  \\
    \bottomrule
    \end{tabular}
  \label{tab:c3sasr:single}%
\end{table}%

\subsubsection{Single Sequence Encoder}
We first evaluate the performance of the single encoders between SASRec and C3SASR model. The item and sequence representation dimensions of C3SASR and SASRec are set to 64, and other settings remain unchanged. The performance is compared at $\lambda = 0$ and 0.5, respectively. $\lambda = 0$ means that the model is optimized for recommendation accuracy only, while 0.5 represents a tradeoff between recommendation accuracy and calibration. The performance comparison is presented in Table \ref{tab:c3sasr:single}, where the best performance is indicated by bolded.

Overall, C3SASR models outperform the SASRec model at both $\lambda$ of 0 and 0.5. Focusing on the model’s accuracy at $\lambda = 0$, it is found that the C3SASR-series models outperforms SASRec on both data sets. For example, on the Ml-1m dataset, the performance of C3CSASR on MRR@10 is 8.25$\%$ higher than that of SASRec. On the Tmall dataset, the recommendation accuracy of the C3SASR series model is also better than that of SASRec. When $\lambda$ = 0.5, C3SASR has higher accuracy than the SASRec, but the calibration performs slightly worse than SASRec. For example, on Tmall, the MR of C3CSASR is 0.0891, higher than 0.0854 of SASRec. However, on the performance of $C_{KL}@10$, the C3CSASR model is 2.2168, while SASRec is 2.1490. The possible reason is that a single sequence encoder facing two optimization targets has a performance bottleneck problem because of the shared underlying parameters. In contrast, under the DACSR framework, the capability of C3SASR is amplified due to the decomposition-aggregation structure of the DACSR framework. Therefore, the performance improvement of the model under the DACSR framework is more prominent.

\subsubsection{Under the DACSR Framework}
In this section, we compare the performance under the DACSR framework for calibrated recommendations. The dimensions of each single encoder of DACSR framework are set to 64. We also involve the original SASRec model and the CaliRec model where the dimensions of items and sequence embeddings are set to 128. Performances are listed in Table \ref{tab:c3sasr:overall}. DACSR (C3) and DACSR (C3C) represent the DACSR model using C3SASR and C3CSASR models as sequence encoders, respectively, while DACSR represents the model using SASRec as sequence encoders.

In general, the performance of C3SASR-series models under the framework of DACSR has improved in the recommended accuracy of the DACSR model with C3SASR-series models as sequence encoders. In contrast the performance of calibration is similar to that of the DACSR model with SASRec as sequence encoders. On the Ml-1m dataset, the DACSR (C3C) model achieves optimal accuracy. For example, Rec@20 is 0.3934, 2.34$\%$ higher than DACSR, and MRR@20 is 0.1392, which is about 4$\%$ higher than DACSR. On the Tmall dataset, DACSR (C3) achieves the best performance. For example, in MRR10 and MRR@20, compared with the DACSR model, the performance of indicators increased by 8.05$\%$ and 7.71$\%$, respectively. Regarding calibration, the performance of C3SASR-series models is similar to that of DACSR with SASRec as the sequence encoder. For example, on the Ml-1m dataset, the CKL@20 of DACSR (C3C) is 0.7113, while DACSR (C3C) is 0.7273, which is close to the 0.7262 of DACSR. It also shows similar performance on Tmall dataset. 

\begin{table}[tbp]
  \centering
  \caption{Performance comparisons under the DACSR framework}
    \begin{tabular}{c|c|ccc|cc}
    \toprule
    Datasets & Metrics & SASRec & CaliRec & DACSR & DACSR(C3) & DACSR(C3C) \\
    \midrule
    \multirow{6}[2]{*}{Ml-1m} & Rec@10 & 0.2627  & 0.2636  & 0.2811  & 0.2780  & \textbf{0.2858 } \\
          & MRR@10 & 0.1203  & 0.1101  & 0.1267  & 0.1279  & \textbf{0.1318 } \\
          & $C_{KL}$@10 & 1.2385  & 0.9722  & 1.0615  & \textbf{1.0492 } & 1.0617  \\
          & Rec@20 & 0.3613  & 0.3616  & 0.3844  & 0.3838  & \textbf{0.3934 } \\
          & MRR@20 & 0.1271  & 0.1168  & 0.1338  & 0.1352  & \textbf{0.1392 } \\
          & $C_{KL}$@20 & 0.8548  & 0.7322  & 0.7262  & \textbf{0.7113 } & 0.7273  \\
    \midrule
    \multirow{6}[2]{*}{Tmall} & Rec@10 & 0.1451  & 0.1464  & 0.1517  & \textbf{0.1561 } & 0.1547  \\
          & MRR@10 & 0.0862  & 0.0846  & 0.0857  & \textbf{0.0926 } & 0.0916  \\
          & $C_{KL}$@10 & 2.5004  & 2.0710  & \textbf{2.0114 } & 2.0496  & 2.0409  \\
          & Rec@20 & 0.1749  & 0.1753  & 0.1855  & \textbf{0.1895 } & 0.1848  \\
          & MRR@20 & 0.0883  & 0.0866  & 0.0881  & \textbf{0.0949 } & 0.0937  \\
          & $C_{KL}$@20 & 2.1103  & 1.7943  & \textbf{1.6240 } & 1.6566  & 1.6487  \\
    \bottomrule
    \end{tabular}%
  \label{tab:c3sasr:overall}%
\end{table}%
The above experiments show that the C3SASR series models improve the characterization ability of sequences and the accuracy of model prediction when obtaining similar calibration. If a more calibrated recommendation list is needed, there is more room for C3SASR-series models. A more calibrated recommendation list can be obtained by sacrificing a sure accuracy.
Therefore, this section explores whether it is possible to obtain a more calibrated recommendation list than DACSR while obtaining a similar accuracy as DACSR by increasing the $\lambda$ in the loss function $L_w$. In this section, $\lambda$ is raised to 0.6 and 0.7, experiments are conducted on the Tmall dataset, the experimental results are provided in Table \ref{tab:c3sasr:lambda-tmall}.
\begin{table}[tbp]
  \centering
  \caption{Performance comparisons when $\lambda$ raises}
    \begin{tabular}{c|c|cccccc}
    \toprule
    Models & $\lambda$ & Rec@10 & MRR@10 & $C_{KL}$@10 & Rec@20 & MRR@20 & $C_{KL}$@20 \\
    \midrule
    DACSR & 0.5   & 0.1517  & 0.0857  & 2.0114  & 0.1855  & 0.0881  & 1.6240  \\
    \midrule
    \multirow{3}[2]{*}{DACSR(C3)} & 0.5   & \textbf{0.1561 } & \textbf{0.0926 } & 2.0496  & \textbf{0.1895 } & \textbf{0.0949 } & 1.6566  \\
          & 0.6   & 0.1529  & 0.0900  & 1.9394  & 0.1852  & 0.0922  & 1.5531  \\
          & 0.7   & 0.1497  & 0.0891  & \textbf{1.8111 } & 0.1799  & 0.0912  & \textbf{1.4061 } \\
    \bottomrule
    \end{tabular}%
  \label{tab:c3sasr:lambda-tmall}%
\end{table}%

As shown in the table, when $\lambda$ is raised to 0.6, the recall performance of DACSR (C3) is equivalent to that of DACSR, while the performance of MRR is better than that of DACSR. For example, the performance of the original DACSR model and DACSR (C3) model on Rec@20 are 0.1855 and 0.1852, while on MRR@20 are 0.0881 and 0.0922, still 4.65$\%$ higher. In terms of calibration of the recommended list, the performance of $C_{KL}@20$ decreases from 1.6240 to 1.5531. DACSR(C3) has better sorting performance and a more accurate recommendation list at $\lambda$ of 0.6 compared to DACSR at $\lambda$ of 0.5. And in $\lambda$= 0.7, the performance of DACSR (C3) in terms of recall rate decreases to a certain extent compared with the original DACSR. However, the sorting performance remains ahead, and the calibration criteria were further improved. The comparative experiment shows that our C3SASR has better sequence expression ability and can better deal with multi-objective optimization tasks.

\subsection{RQ2: Parameter Influence}
This section concerns the effect of convolution kernel size on the model in the light causal convolution operation. The convolution kernel size represents the number of local items considered. The convolution kernel's default size in earlier studies was 3. This section varies its value from 1 to 5 and tracks how the performance changes. The results are presented in Table \ref{tab:c3sasr:kernel-tmall}. The results reveal that the network performs best with a convolutional kernel size of 3. Larger or smaller values reduce the performance accordingly. It is because the current item is associated with only some preorder items. A smaller value does not capture enough contextual information, whereas a larger value introduces irrelevant aspects.
\begin{table}[tbp]
  \centering
  \caption{Performance comparisons when $k$ changes}
    \begin{tabular}{c|cccccc}
    \toprule
    kernel & Rec@10 & MRR@10 & $C_{KL}$@10 & Rec@20 & MRR@20 & $C_{KL}$@20 \\
    \midrule
    1     & \textbf{0.1561 } & 0.0911  & 2.0411  & 0.1875  & 0.0932  & 1.6572  \\
    2     & 0.1511  & 0.0906  & \textbf{2.0329 } & 0.1827  & 0.0928  & \textbf{1.6401 } \\
    3     & \textbf{0.1561 } & \textbf{0.0926 } & 2.0496  & \textbf{0.1895 } & \textbf{0.0949 } & 1.6566  \\
    4     & 0.1548  & 0.0908  & 2.0461  & 0.1885  & 0.0931  & 1.6634  \\
    5     & 0.1544  & 0.0910  & 2.0387  & 0.1870  & 0.0933  & 1.6438  \\
    \bottomrule
    \end{tabular}%
  \label{tab:c3sasr:kernel-tmall}%
\end{table}%

We also analyze the effect of changes in the values of the parameters $k_2$ and $\alpha$ on the changes in the model. C3CSASR uses a light causal convolution mechanism in the sequence representation generation, and $k_2$ is its convolution kernel size, just the number of local items considered. Moreover, $\alpha$ represents the weight of the representation obtained by transforming the last item $x_t$ of the sequence, and 1-$\alpha$ represents the weight occupied by the augmented representation obtained from the local items. $k_2$ is set to 3, and $\alpha$ is set to 0.8 by default for C3CSASR. This section takes its $k_2$ to an additional value of 5 and $\alpha$ to an additional value of 0.5. Performance comparison is presented in Table \ref{tab:c3sasr:k2alpha}, where "None" represents C3SASR without the local enhancement strategy.
On the Ml-1m dataset, increasing $k_2$ or decreasing $\alpha$ causes a decrease in model performance. It indicates that too much consideration of the local information of the series tends to introduce noise, resulting in a performance decrease. In contrast, introducing local information enhancement appropriately leads to increased model performance. However, C3SASR slightly outperforms C3CSASR on the Tmall dataset. One possible reason is that the Tmall dataset has a greater difficulty predicting the following behavior than the Ml-1m dataset. In this case, the mechanism of introducing local enhancement in generating sequence representations may not positively affect the model performance.
\begin{table}[tbp]
  \centering
  \caption{Performance comparisons under different value of $k_2$ and $\alpha$.}
    \begin{tabular}{c|c|c|cccccc}
    \toprule
    Dataset & $k_2$    & $\alpha$ & Rec@10 & MRR@10 & $C_{KL}$@10 & Rec@20 & MRR@20 & $C_{KL}$@20 \\
    \midrule
    \multirow{4}[2]{*}{Ml-1m} & None  & None  & 0.2780  & 0.1279  & 1.0492  & 0.3838  & 0.1352  & 0.7113  \\
          & 3     & 0.8   & 0.2858  & \textbf{0.1318 } & 1.0617  & \textbf{0.3934 } & \textbf{0.1392 } & 0.7273  \\
          & 3     & 0.5   & 0.2762  & 0.1280  & \textbf{1.0368 } & 0.3876  & 0.1357  & \textbf{0.6938 } \\
          & 5     & 0.8   & \textbf{0.2866 } & 0.1300  & 1.0649  & 0.3884  & 0.1370  & 0.7292  \\
    \midrule
    \multirow{4}[2]{*}{Tmall} & None  & None  & \textbf{0.1561 } & \textbf{0.0926 } & 2.0496  & \textbf{0.1895 } & \textbf{0.0949 } & 1.6566  \\
          & 3     & 0.8   & 0.1547  & 0.0916  & \textbf{2.0409 } & 0.1848  & 0.0937  & \textbf{1.6487 } \\
          & 3     & 0.5   & 0.1558  & 0.0918  & 2.0564  & 0.1877  & 0.0940  & 1.6586  \\
          & 5     & 0.8   & 0.1555  & 0.0917  & 2.0475  & 0.1865  & 0.0938  & 1.6553  \\
    \bottomrule
    \end{tabular}%
  \label{tab:c3sasr:k2alpha}%
\end{table}%
\subsection{Ablation Study}
\begin{table}[tbp]
  \centering
  \caption{Performance comparisons among variants.}
    \begin{tabular}{c|cccccc}
    \toprule
    Models & Rec@10 & MRR@10 & $C_{KL}$@10 & Rec@20 & MRR@20 & $C_{KL}$@20 \\
    \midrule
    DACSR & 0.1517  & 0.0857  & 2.0114  & 0.1855  & 0.0881  & 1.6240  \\
    FFN & 0.1537  & 0.0902  & 2.0091  & 0.1860  & 0.0924  & 1.6228  \\
    w/o CC & 0.1498  & 0.0887  & \textbf{1.9981 } & 0.1807  & 0.0909  & \textbf{1.6031 } \\
    RawConv & 0.1560  & 0.0898  & 2.0502  & 0.1886  & 0.0921  & 1.6597  \\
    DACSR(C3) & \textbf{0.1561 } & \textbf{0.0926 } & 2.0496  & \textbf{0.1895 } & \textbf{0.0949 } & 1.6566  \\
    \bottomrule
    \end{tabular}%
  \label{tab:c3sasr:ablation-tmall}%
\end{table}%
This section focuses on the impact of each module of the model on performance by comparing some modules of C3SASR. Specifically, this section focuses on the following variants of the model:
\begin{itemize}
\item FFN: the linear layer of the SASRec model is employed in the sequence representation generation layer, replacing the cheap convolution mechanism employed by C3SASR.
\item w/o CC: the C3 mechanism in the self-attentive layer is removed.
\item RawConv: the C3 mechanism employs the original one-dimensional convolution layer. 
\end{itemize}
The experimental setup is kept constant, and the results are presented in Table \ref{tab:c3sasr:ablation-tmall}, where the best performance is indicated using bolded.

The FFN model removes the cheap convolution mechanism of the list feature generation layer and uses only linear layers. In comparison with the full C3SASR model, its accuracy decreases relatively. For example, Rec@20 decreases from 0.1895 to 0.1860, and MRR@20 decreases from 0.0949 to 0.0924, while the calibration performance is slightly improved. It indicates that using a cheap convolution mechanism in the list characterization generation layer helps obtain a better item characterization.
Removing the C3 mechanism in the self-attentive layer causes a decrease in recommendation accuracy. The performance of Rec@20 decreases from 0.1895 to 0.1807, a decrease of 4.64$\%$, and the performance of MRR@20 decreases by 4.21$\%$. Such results indicate that introducing item context information does help to obtain better item representations. Using the original convolution instead of the cheap convolution mechanism proposed in this paper in fusing the item contexts also caused a decrease in model performance. While the performance of recall is similar, the performance of the ranking metric MRR shows a dip. The calibration still performs similarly to the entire model. The comparison shows that our cheap convolution mechanism can obtain more effective item representations with fewer parameters, reducing the redundant information present in the original convolution.

\section{Discussion}
In this paper, we proposed a cheap causal convolution mechanism which aims to 
incorporate local information and reduce the redundant information. Experiments on benchmark datasets validated the effectiveness of proposed model. Experiments under different settings of parameters indicate that an appropriate range of local context information is useful to improve the item representation. Meanwhile, the cheap convolutions can improve the embeddings via a lightweight structure. Ablation studies also verified the effectiveness of proposed causal convolutions and cheap convolutions.

However, there are some limitations of our work as below:
\begin{itemize}
    \item The local context information of items are aggregated by kernels with fixed size. However, items may be related to different numbers of pre-order items.
    \item The proposed model aggregated the local information to enhance the item and sequence representation. This increased the time complexity of the model although cheap convolution reduced parameter amount to a certain degree.
\end{itemize}

\section{Conclusion}
In this paper, we construct an improved self-attentive model for sequential recommendation. We use the causal convolution mechanism to fuse the contextual information of items as the input of the self-attentive mechanism and the way of sequence representation generation. Then, to address the problem of redundant information and large parameter size in the convolutional and linear layer, we use a cheap convolutional mechanism to provide effective representations with fewer parameters by reducing the number of channels and enhancing representation. Experimental results on standard datasets show that our model can improve the model's ability to represent sequences and better provide accurate and calibrated recommendation lists.

For the future work, we first aim to design structures for aggregating context information with flexible sizes. In addition, we are also interested in the lighter structure to aggregate local information.

\section*{Author Contributions}
Author J.C. contributes to conceptualization, methodology, software, validation, formal analysis, investigation, resources, data curation, writing-original draft, writing-review and editing of the paper. Author W.W. contributes to conceptualization, methodology, formal analysis, investigation, writing-Original Draft, writing-review and editing, supervision of the paper. Author L.S. contributes to conceptualization, writing-review and editing of this paper. Authors Y.J., W.H., X.C. and W.Z. contribute to writing-review and editing of the paper. Author L.H. contributes to writing-review and editing and supervision of the paper. 

\section*{Data availability}
We conducted experiments on two benchmark datasets: MovieLens 1M (link: https://grouplens.org/datasets/movielens/1m, accessed at 14 Sept 2021), and Tmall (link:https://tianchi.aliyun.com/dataset/dataDetail?dataId=53, accessed at 3 Jan 2022)

\section*{Acknowledgements}
This work is funded by National Natural Science Foundation of China (under project No. 61907016), Science and Technology Commission of Shanghai Municipality, China (under project No. 21511100302), and Natural Science Foundation of Shanghai (under project No. 22ZR1419000). It is also supported by The Research Project of Shanghai Science and Technology Commission (20dz2260300) and The Fundamental Research Funds for the Central Universities.

\section*{Competing Interests}
All authors certify that they have no affiliations with or involvement in any organization or entity with any financial interest or non-financial interest in the subject matter or materials discussed in this manuscript.

\end{document}